# Spectral description of fluctuating electromagnetic fields of solids including a self consistency

Illarion Dorofeyev

Institute for Physics of Microstructures, Russian Academy of Sciences,

603950, GSP-105 Nizhny Novgorod, Russia

Phone:+7 910 7934517, Fax:+7 8312 675553

E-mail1: Illarion1955@mail.ru , E-mail2: dorof@ipm.sci-nnov.ru

## Abstract

Another way to evaluate the spectral-correlation properties of thermal fields of solids is suggested. Such a method takes into account detailed structure of the interface transition layer separating one bulk region from those of the vacuum region. The principal element here is the surface linear response function of an arbitrary inhomogeneous electron subsystem of solids. Along with straightforward using the known response functions, for example in a hydrodynamic approach, the suggested method allows calculating the response functions self-consistently based on the time dependent density functional theory. The self-consistent calculation of the linear response function followed by an application of the fluctuation-dissipation theorem yield in the spectral power densities of the fluctuating charge and current correlation functions. The final step of this sequence of actions aiming in obtaining the spectral-correlation properties of thermal fields of solids at any distances.

PACS indexing codes: 05.40.-a; 02.50.Ey; 05.10.Gg; 05.70.Np;

## I. INTRODUCTION

Theory of fluctuating electromagnetic (FEM) fields of solids is grounded on the Maxwell's equations with the fluctuation-dissipation theorem (FDT) [1-8]. The key point of this theory is to solve the boundary-value problem of the phenomenological electrodynamics. Knowledge of the Green's tensor of the boundary-value problem together with the FDT enables us to evaluate the spectral and correlation properties of thermal fields of solids [9-16]. It should be emphasized that in accordance with such formulation of the problem, it is clearly that the interface is simply an abstract mathematical line between of material domains. It is true despite the local or nonlocal description of optical properties. At the same time, an adequate calculation of subsurface physical and chemical processes takes most detailed description of an interface as some transition layer between of materials with a variable chemical and structural composition. Obviously, it can be achieved only by quantum mechanical means. The interface transition layer consists the electron subsystem, the nuclear lattice (regular or not) subsystem and the regular and fluctuating electromagnetic fields. We are dealing with here with thermally stimulating fluctuating electromagnetic fields originating from charge fluctuations within the material domain. It is well known that the near-field zone in the heterogeneous system "solid-vacuum" is composed mostly by so-called the evanescent electromagnetic waves. The spectra of thermal electromagnetic fields consist of series of resonances relating to the surface electromagnetic waves (surface polaritons) within the spectral ranges where such surface eigenstates can be excited.

It should be noted that the surface electromagnetic states have an important role in different physical processes including those in the van der Waals interaction of bodies, in the heat transfer between the bodies at small distances, in the continuous growth of crystals, in the Raman-scattering characteristics, in the capture of atoms, molecules, and coherent material states, in photochemistry, in surface phenomena such as the adsorption and desorption phenomena, the heterogeneous chemical catalysis, and etc. In practice, surface polaritons can be excited by laser radiation and by a beam of particles, or by internal thermal fluctuations inside a body. Thermally excited electromagnetic fields within a body due to charge and currents fluctuations partially reflect on the vacuum-sample interface return back to the body, and in part the fields penetrate into a vacuum region outside the body, where the penetrating waves form the electromagnetic background in the near- and far-field regimes. Properties of the propagating and evanescent electromagnetic fluctuating fields are crucially different [9-16]. In the near field the energy density can be much larger in magnitude than in the far field. The matter is that the optical properties and sample geometry have a strong influence on the characteristics of thermally excited near fields. As a result, the noise spectra in the near field regime differ essentially from the noise spectra in the far field regime. Moreover, the coherence properties of thermal electromagnetic fluctuations in a near field regime are extremely different from those of the propagating waves. The above described results were obtained within the framework of the electrodynamics of continuous media.

It goes without saying that the phenomenological approach is valid only at comparatively large distances from solids, at least larger than the interatomic distance in the sample lattice. Along with, we have another way to evaluate the spectral-correlation properties of thermal fields of solids omitting the theory developed in [1-8], and (in addition to the complete description of thermal fields) it is possible to determine in detail the properties of the interface transition layer separating one bulk region from those of the vacuum region. The principal foundation here is the static version of the density functional theory (DFT) [17, 18] for zero and nonzero [19] temperatures, and the time dependent density functional theory (TDDFT) [20-25]. TDDFT allows calculating the linear response function of an arbitrary inhomogeneous electron system of solids followed by an application of the fluctuation-dissipation theorem in order to obtain the spectral power densities of the charge and current correlation functions. In the final step of this scenario, the spectral-correlation properties of thermal fields of solids at any distances (including those within the solids) will be calculated.

The paper is organized as follows. In Sec.II we provide theoretical basis to calculate the spectral-correlation characteristics of thermal fields using TDDFT. Analytical formulas for the spectral power densities of the components of thermal fields in the partial case of a linear response function for a half-space in the hydrodynamic approach and corresponding illustrations are presented in Sec.III, subsection 1. Numerical results for distance dependence of the spectral power densities of thermal fields in case of the self-consistent approach comparing with the phenomenological approach are demonstrated in Sec.III, subsection 2. Our conclusions are given in Sec.IV. Finally, in appendixes we provide some accepted definitions and details of the self-consistent solution of the Kohn-Sham system of equations.

## II. FORMALISM

Here, we formulate the problem of calculating the spectral-correlation tensors of thermal fields via the linear response functions of spatially inhomogeneous systems. Phenomenological theory adequately describes the fluctuating electromagnetic fields at comparatively large distances from the surface of a body. That is why; we restrict our consideration to the subsurface domains at distances much smaller than the typical wavelengths of thermal spectra of solids. Thus, we consider the quasistationary (evanescent) fields. In general, the heated body of interest has an arbitrary form and volume $V$. The sources of thermal fields are the fluctuating charges $\rho(\vec{r},t)$ and currents $\vec{j}(\vec{r},t)$ within this volume. The formulas for the field's components are as follows

$$E_i(\vec{r},t) = -\int_V d^3r' \rho(\vec{r}',t) \frac{\partial}{\partial x_i} \frac{1}{|\vec{r}-\vec{r}'|}, \qquad H_i(\vec{r},t) = \frac{1}{c}\int_V d^3r' e_{ik\ell} \frac{\partial}{\partial x_k} \frac{j_\ell(\vec{r}',t)}{|\vec{r}-\vec{r}'|}, \qquad (1)$$

where $e_{ik\ell}$ is the totally antisymmetric unit pseudotensor of a third rank. Corresponding symmetrized correlation tensors of thermal fields are obtained by composing the components from Eqs.(1), for example,

$$\langle E_i(\vec{r}_1,t) E_k(\vec{r}_2,t+\tau) \rangle^S = (1/2)\{\langle E_i(\vec{r}_1,t) E_k(\vec{r}_2,t+\tau) + E_k(\vec{r}_2,t+\tau) E_i(\vec{r}_1,t) \rangle\}, \qquad (2)$$

and then we have

$$\langle E_i(\vec{r}_1,t) E_k(\vec{r}_2,t+\tau) \rangle^S = \iint_V d^3r' d^3r'' \langle \rho(\vec{r}',t)\rho(\vec{r}'',t+\tau) \rangle^S \frac{\partial}{\partial x_i} \frac{1}{|\vec{r}_1-\vec{r}'|} \frac{\partial}{\partial x_k} \frac{1}{|\vec{r}_2-\vec{r}''|}. \qquad (3)$$

Writing the symmetrized functions in Eq.(3) in forms as follows

$$\langle E_i(\vec{r}_1,t) E_k(\vec{r}_2,t+\tau) \rangle^S = \int_{-\infty}^{\infty} \frac{d\omega}{2\pi} g_{ik}^{(E)}(\vec{r}_1,\vec{r}_2;\omega) \exp(-i\omega\tau), \qquad (4)$$

and

$$\langle \rho(\vec{r}',t)\rho(\vec{r}'',t+\tau) \rangle^S = \int_{-\infty}^{\infty} \frac{d\omega}{2\pi} (\rho(\vec{r}')\rho(\vec{r}''))_\omega \exp(-i\omega\tau), \qquad (5)$$

we have for the spectral power densities

$$g_{ik}^{(E)}(\vec{r}_1,\vec{r}_2;\omega) = \iint_V d^3r' d^3r'' (\rho(\vec{r}')\rho(\vec{r}''))_\omega \frac{\partial}{\partial x_i} \frac{1}{|\vec{r}_1-\vec{r}'|} \frac{\partial}{\partial x_k} \frac{1}{|\vec{r}_2-\vec{r}''|}, \qquad (6)$$

$$g_{ik}^{(H)}(\vec{r}_1,\vec{r}_2;\omega) = \frac{1}{c^2} \iint_V d^3r' d^3r'' e_{is\ell} e_{kmn} (j_\ell(\vec{r}') j_n(\vec{r}''))_\omega \frac{\partial}{\partial x_s} \frac{1}{|\vec{r}_1-\vec{r}'|} \frac{\partial}{\partial x_m} \frac{1}{|\vec{r}_2-\vec{r}''|}, \qquad (7)$$

$$g_{ik}^{(HE)}(\vec{r}_1,\vec{r}_2;\omega) = -\frac{1}{c} \iint_V d^3r' d^3r'' e_{is\ell} \frac{\partial}{\partial x_s} \frac{(j_\ell(\vec{r}')\rho(\vec{r}''))_\omega}{|\vec{r}_1-\vec{r}'|} \frac{\partial}{\partial x_k} \frac{1}{|\vec{r}_2-\vec{r}''|}, \qquad (8)$$

where $(\rho(\vec{r}')\rho(\vec{r}''))_\omega$, $(j_\ell(\vec{r}') j_n(\vec{r}''))_\omega$ and $(j_\ell(\vec{r}')\rho(\vec{r}''))_\omega$ are the symmetrized spectral power densities for the fluctuating charges, currents and crossed characteristics, see Appendix A. Then, taking into account FDT from Eq.(A5) corresponding to charges fluctuations $(\rho(\vec{r}')\rho(\vec{r}''))_\omega = \hbar Coth(\hbar\omega/2k_B T) \text{Im}\{\chi(\vec{r}',\vec{r}'';\omega)\}$, it is directly follows from Eq.(6), for instance, the spectral power density of electric fluctuations

$$g_{ik}^{(E)}(\vec{r}_1,\vec{r}_2;\omega) = \hbar Coth(\hbar\omega/2k_B T) \iint_V d^3r' d^3r'' \text{Im}\{\chi(\vec{r}',\vec{r}'';\omega)\} \frac{\partial}{\partial x_i} \frac{1}{|\vec{r}_1-\vec{r}'|} \frac{\partial}{\partial x_k} \frac{1}{|\vec{r}_2-\vec{r}''|}, \qquad (9)$$

where the linear response function $\chi(\vec{r}',\vec{r}'';\omega)$, in its turn, is defined by the following relation between the induced charge $\rho_{ind}$ and external potential $\phi_{ext}$

$$\rho_{ind}(\vec{r},\omega) = \int_V d^3r' \, \chi(\vec{r},\vec{r}';\omega) \phi_{ext}(\vec{r}',\omega). \qquad (10)$$

The appearing charge fluctuation $\rho = \rho_{ind} - \bar{\rho}$, where $\bar{\rho}$ is the mean charge within a system, which is indicated by a zero value.

The main part of the proposed method used to calculate the spectral and correlation properties of thermal fields of solids is concerned with the time dependent density functional theory [20-25]. In a framework of this method the linear response of an arbitrary spatially bounded system of volume $V$ is a solution of the integral equation

$$\chi(\vec{r}',\vec{r}'';\omega) = \chi^0(\vec{r}',\vec{r}'';\omega) + \iint_V d^3r_1 d^3r_2 \, \chi^0(\vec{r}',\vec{r}_1;\omega) V(\vec{r}_1,\vec{r}_2;\omega) \chi(\vec{r}_2,\vec{r}'';\omega), \qquad (11)$$

where $\chi^0(\vec{r}',\vec{r}'';\omega)$ is the reduced linear response of the system of noninteracting electronic states, $V(\vec{r}_1,\vec{r}_2;\omega)$ is the electron-electron interaction potential, including the Coulomb and exchange-correlation parts, in general.

Corresponding literature for this topic is rather vast, but, for the illustration purpose of this paper we further deal with a problem as applied to a half-space and metallic films in the so-called jellium model. It should be emphasized that within a framework of the self-consistent Kohn-Sham scheme a structure of the boundary transition layer can be described in detail, differing to the phenomenological approach. In our case this is a spatial distribution of electrons nearby an interface. The Kohn-Sham system of equations and calculated electron density nearby aluminum sample are done in appendix C. We note that in case of a half-space and films the linear response of the system of free electronic states $\chi^0$ is known analytically. But, the analytical expression for $\chi^0$ not contain a finite value of dissipation in a system under study. Instead of this an infinitesimal adiabatic parameter is included in a denominator of $\chi^0$. For reasons of necessity and practicability this parameter is supposed by a finite value, but, this is not correct way because of a violation of the charge conservation law. Now, this is the main disadvantage of the self-consistent calculations based on Eq.(11) for our purposes.

### III. NUMERICAL EXAMPLES OF SOLUTIONS AND DISCUSSIONS
#### 1. In the framework of the hydrodynamic approach

Of course, the most appropriate calculation of the linear response function is based on the self-consistent approach leading to Eq.(11). But, from one side, it takes a lot of computer time, and from other side, a finite rate of dissipation in $\chi^0$ is necessary from physical point of view, due to the fluctuation dissipation theorem. In this paragraph we use nonself-consistent, but a more simple expression for the linear response function in Eq.(9) for a half-space known from [27] in the hydrodynamical approach. Arguments in favor of this choice are connected with taking into account of a finite value of dissipation in $\chi$ and relative simplicity. A geometry of the problem is as follows: a half-space within a spatial domain $z \leq 0$ is filled by a metal contacting with a vacuum in $z > 0$. Spectral power densities of thermal fields are calculated at the distance $h$ from a metallic half-space.

Substituting a formula for $\chi$ from the work [27] to Eq.(9), see appendix B, resulting in the expression for SPD of the z-component of the fluctuating field

$$g_{zz}^{(E)}(h,\omega) = \frac{\hbar \omega_P^2}{\beta^2[\exp(\hbar\omega/k_B T)-1]} \text{Im} \int_0^\infty dQ \exp(-2Qh) \frac{Q}{2Q_L}\left[1 - \frac{Q_L - Q}{Q_L + Q}(1+\gamma)\right], \quad (12)$$

where all parameters are given in appendix B.

Herewith, $g_{zz}^{(E)} = 2g_{xx,yy}^{(E)}$. Figure 1 exemplifies the normalized spectral power density of z-component of the electric thermal field $\tilde{g}_{zz}^{(E)}(h)$ versus a distance $h$ in accordance with Eq.(12) (red curve), and in accordance with a local phenomenological theory of thermally stimulated fields [1-3,10-14] (black line) at the fixed frequency $\omega = 0.1\omega_P$. The functions are normalized at the value $g_{zz}^{(E)}(h=0)$. The quantity $h=0$ is fixed at the edge of the positive background of a metal. In calculations, the parameters of $\omega_P = 2.3 \times 10^{16} \, rad/s$, $\nu = 1.3 \times 10^{14} \, s^{-1}$ [28], $v_F = 2.03 \times 10^8 \, cm/s$ were used, corresponding to aluminum. It is seen from this figure that SPD fully saturates at small distances up to interatomic scales.

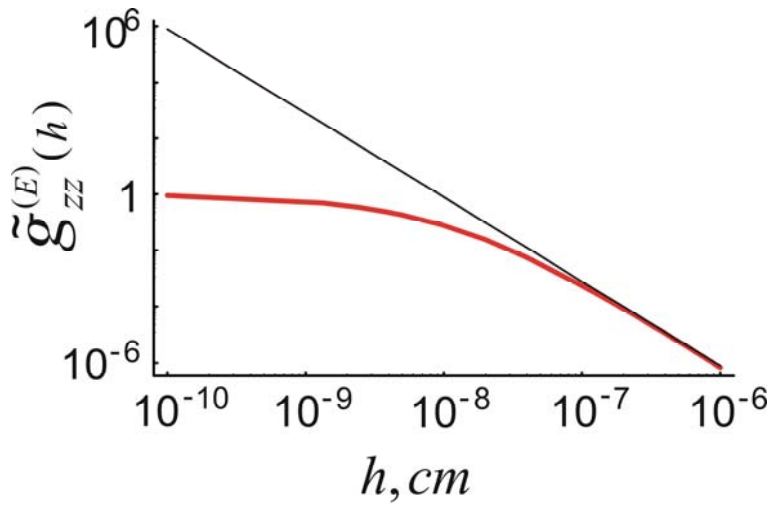

Fig. 1

Moreover, the suggested approach allows calculating spectral and correlation properties of thermal fields within a sample at any position in a transition layer. It should be noted that the larger $h$, the closer the red curve to the phenomenological black line up to a merging.

In principal, in the framework of phenomenological theory it is possible to take into account of nonlocal optical effects yielding to a saturation of the spectral power densities of fields at $h \to 0$, [2, 30, 31]. But, in any case (local or nonlocal) we have to solve corresponding boundary-value electrodynamical problem without of detailed description the transition layer formed by electrons both within a metal and vacuum domains. Obviously, it is possible only on a basis of a physically relevant quantum theory, for example, like the TDDFT.

Taking into account a possibility of the correct calculations of SPD within a transition layer, an agreement between properties of fields inside of a bulk material and in a vacuum domain is worth studied. It is well known the spectral densities of fluctuations of electromagnetic fields in an infinite, homogeneous, spatially nonlocal and lossy medium, see, for instance [2, 3]. For nonmagnetic media, when $\varepsilon^{tr} = \varepsilon^{\ell} = \varepsilon(\omega,k) = \varepsilon'(\omega,k) + i\varepsilon''(\omega,k)$ the trace of spectral densities of electric components are expressed as follows

$$g_{EE}(\omega,k) = \frac{\theta(\omega,T)\omega^3}{\pi^3} \frac{\varepsilon''(\omega,k)}{\left|k^2c^2 - \omega^2\varepsilon(\omega,k)\right|^2} + \frac{\theta(\omega,T)}{2\pi^3\omega} \frac{\varepsilon''(\omega,k)}{\left|\varepsilon(\omega,k)\right|^2}, \qquad (13)$$

where $g_{EE}(\omega,k) = Tr\{g^{(E)}_{\alpha\beta}(\omega,k)\}$ is the sum of SPD of all components.

Taking the inverse transform in case of isotropic space

$$g_{EE}(\omega) = 4\pi \int_0^\infty g_{EE}(\omega,k) k^2 dk, \qquad (14)$$

we compare this quantity within a medium with $g_{EE}(\omega) = 2g^{(E)}_{zz}(\omega)$ in a vacuum in accordance with Eq.(12). In calculations we use $\varepsilon(\omega,k) = 1 - \omega_P^2/(\omega^2 + iv\omega - \beta^2k^2)$ with same parameters as above. Fig.2 shows the normalized value $g_{EE}(\omega)/u_{BB}(\omega)$ within a metal at the left panel and $g_{EE}(\omega,h)/u_{BB}(\omega)$ in a vacuum at the right panel of the figure at different fixed frequencies in accordance with Eq.(14) and (12), correspondingly. Normalization was done to the spectral energy density $u_{BB}(\omega)$ of the black body radiation in vacuum in accordance with the Planck law.

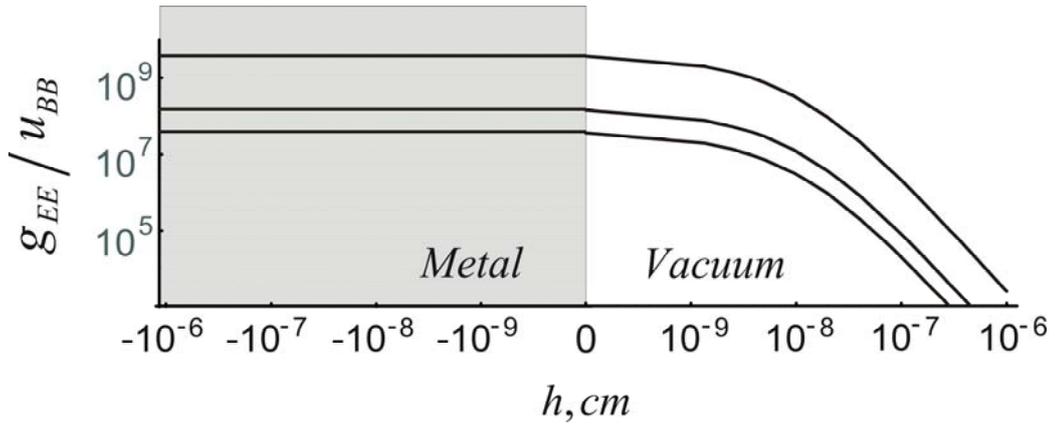

Fig.2

It is clearly seen the excellent coincidence of SPD just at the interface between two spatial domains.

2. **The inclusion of self-consistency**

The suggested method allows us to calculate spectral correlation properties of thermal fields both in self-consistent and in nonself-consistent approaches, as it was described in above paragraph. To exemplify the self-consistent calculations we considered a plane-parallel film of aluminum ($r_s = 2.07$) of a fixed thickness ($d = 4 \times 10^{-7} cm$). From one side a finite thickness of sample was chosen due to discreetness of energy levels to simplify numerical calculations, and from other side because of opportunities of our personal computer. Our main goal here to demonstrate the approach, that is why we restrict our calculations only by frameworks of the random-phase approximation taking into account the Hartree term in an interaction energy in Eq.(C7). In this case we use the Fourier transforms $\chi(z_1, z_2; Q; \omega)$ from Eq.(B1) instead of Eq.(11)

$$\chi(z', z''; Q, \omega) = \chi^0(z', z''; Q, \omega) + \int_0^d dz_1 \int_0^d dz_2\, \chi^0(z', z_1; Q, \omega) V(z_1, z_2; Q) \chi(z_2, z''; Q, \omega), \quad (15)$$

where $V(z_1, z_2; Q) = 2\pi \exp(-Q|z_1 - z_2|)/Q$ is the 2D Fourier transform of the function $|\vec{r}_1 - \vec{r}_2|^{-1}$.

Taking into account Eqs.(9), (B1), (C13) we obtain the self-consistent expression for the spectral power density, for instance, for the lateral components of electric fields

$$g^{(E)}_{xx, yy}(h, \omega) = \frac{2\pi \hbar}{\exp(\hbar \omega / k_B T) - 1} \operatorname{Im}\left\{\int_0^\infty dQ\, Q \exp(-2Qh)\left[G^0(Q, \omega) + G(Q, \omega)\right]\right\}, \quad (16)$$

where

$$G^0(Q,\omega) = \sum_{\ell=1}^{\ell_M} \sum_{\ell'=1}^{\ell'_M} F_{\ell\ell'}(Q,\omega) \int_0^d \int_0^d dz'dz'' \exp[-Q(z'+z'')]\phi_\ell(z')\phi_{\ell'}(z')\phi_\ell(z'')\phi_{\ell'}(z''), \quad (17)$$

$$G(Q,\omega) = \int_0^d \int_0^d dz'dz'' \exp[-Q(z'+z'')] \int_0^d \int_0^d dz_1 dz_2 \, \chi^0(z',z_1;Q,\omega)V(z_1,z_2;Q)\chi(z_2,z'';Q,\omega). (18)$$

For chosen form of $\phi_\ell(z)$ in Eq.(C11) we have after elementary integration in Eq.(17)

$$G^0(Q,\omega) = \frac{4}{d^2} \sum_{\ell=1}^{\ell_M} \sum_{\ell'=1}^{\ell'_M} F_{\ell\ell'}(Q,\omega) \sum_{s=1}^{\infty} \sum_{s'=1}^{\infty} b_s^{(\ell)} b_{s'}^{(\ell')} \left( \frac{2\pi^2 Q d^2 [1-(-1)^{s+s'} \exp(-Qd)]ss'}{\pi^4(s^2-s'^2)^2 + 2\pi^2 Q^2 d^2(s^2+s'^2) + Q^4 d^4} \right)^2, \quad (19)$$

where coefficients $b_s^{(\ell)}$ are sought for self-consistently in appendix C.

It should be noted that $g_{zz}^{(E)} = 2 g_{xx,yy}^{(E)}$.

We calculated of $g_{xx,yy}^{(E)}$ in Eq.(14) by iterations, keeping only the term $G^0$ in this equation as the zero iteration, then putting $\chi = \chi^0$ in Eq.(16) for the first iteration, and so on.

In case of so-called infinite barrier model (IBM) for free electrons, $b_s^{(\ell)} = \delta_{\ell s}$ in Eq.(19), where $\delta_{\ell s}$ is the Kronecker symbol. The functions in Eq.(17) are simply $\phi_\ell(z) = (2/d)^{1/2} \sin(\ell\pi z/d)$ in this case, and we have

$$G^0(Q,\omega) = \frac{4}{d^2} \sum_{\ell=1}^{\ell_M} \sum_{\ell'=1}^{\ell'_M} F_{\ell\ell'}(Q,\omega) \left( \frac{2\pi^2 Q d^2 [1-(-1)^{\ell+\ell'} \exp(-Qd)]\ell\ell'}{\pi^4(\ell^2-\ell'^2)^2 + 2\pi^2 Q^2 d^2(\ell^2+\ell'^2) + Q^4 d^4} \right)^2. \quad (20)$$

In our calculations the parameter $\eta$ in Eq.(C14) was considered as a finite number $\nu = 1.3 \times 10^{14} s^{-1}$ as well as in the hydrodynamic model. Fig.3 shows the distance dependence of the normalized spectral power density $g_{zz}^{(E)} = 2 g_{xx,yy}^{(E)}$. The red curve 1 corresponds to the case of so-called infinite barrier model (IBM) when we hold in Eq.(16) only one term $G^0(Q,\omega)$ from Eq.(20). The black line exemplifies the same SPD distance dependence based on a local phenomenological theory of thermally stimulated fields [1-3, 10-14]. The case of self-consistent calculations of SPD with use of both terms $G^0(Q,\omega)$ and $G(Q,\omega)$ in Eq.(16) in the so-called random phase approach (RPA) is demonstrated by the red curve 2. All curves are normalized at $g_{zz}^{(E)}(h=0)$ of the second red curve. It is clearly seen from this figure that the accounting for the Coulomb interaction and self-consistency yield to the crucial decrease of SPD as compare with the IBM model of free noninteracting electrons. Herewith, the curve 2 asymptotically closer to the black line than the curve 1 at large distances.

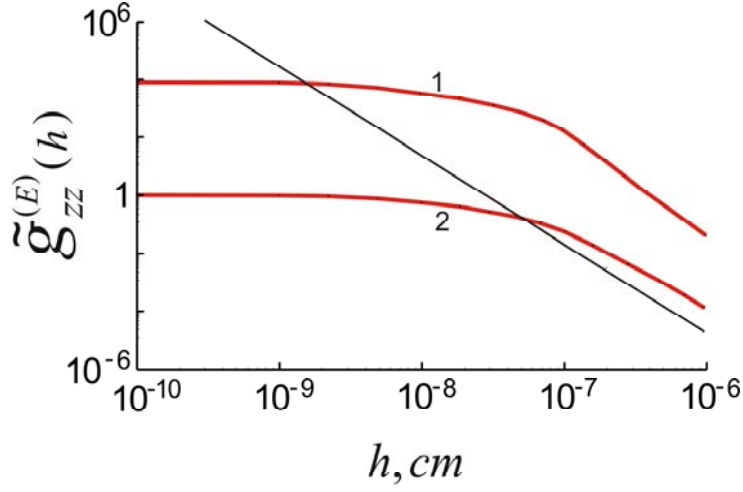

Fig.3

Perhaps the deviation of the curve 2 from the black line at large distances is connected with the noncorrect consideration of the adiabatic parameter $\eta \to 0$ as a finite number $\nu$, despite of the condition $\nu \ll \omega_p$. A correct expression for the reduced linear response $\chi^0$ of the Kohn-Sham states in Eqs.(15), (C13) with finite value of dissipative losses is a matter of a special theoretical study, which is out of our aim in this paper. Another admissible reason of the deviation is connected with an incomplete accounting for all possible interactions in $V$ in Eqs.(11), (15) and $V_{eff}$ in Eq.(C6).

It is quite plausible that accounting for the exchange correlation terms in the interaction potential and using correct reduced linear response $\chi^0$ with finite dissipative losses will give much better coincidence with the phenomenological results at large distances from a sample.
Finally, it should be emphasized that in the framework of the above described method all values of SPD $g_{ik}^{(E)}$ and corresponding spectral energy of fluctuating electromagnetic fields just at the sample surface ($h = 0$) are can be easily calculated.

## IV. CONCLUSION

In this paper we demonstrated another way to evaluate the spectral-correlation properties of thermal fields of solids. Such a way takes into account detailed structure of the interface transition layer separating one bulk region from those of the vacuum region. The principal element here is the surface linear response function of an arbitrary inhomogeneous electron subsystem of solids. Along with straightforward using the known response functions, for example in a hydrodynamic approach, the suggested method allows calculating the response functions self-consistently based on the time dependent density functional theory. The self-consistent calculation of the linear response function followed by an application of the fluctuation-dissipation theorem yield in the spectral power densities of the charge and current correlation functions. The final step of this sequence of actions aiming in obtaining the spectral-correlation properties of thermal fields of solids at any distances, including those within the solids. It should be noted that the well-known phenomenological approach is valid only at comparatively large distances from solids, at least larger than the interatomic distance in the sample lattice. From my point of view, this is a visual demonstration of a possibility of deriving properties of a spatially inhomogeneous bosonic system based on the properties of a spatially inhomogeneous fermionic system.

## ACKNOWLEDGEMENT

Part of this work was performed during the preparation of the project №13-02-97034 - Povolzhie.

## APPENDIX A: FOURIER TRANSFORMS USED AND CHARGE, CURRENT AND CROSSED SPECTRAL POWER DENSITIES

By definition we used the following Fourier transforms for the cross-correlated tensor of two stochastic functions $x_i(t)$ and $x_k(t)$

$$\langle x_i(\vec{r}_1,t) x_k(\vec{r}_2,t')\rangle = \int_{-\infty}^{\infty} \frac{d\omega}{2\pi} \langle x_i(\vec{r}_1) x_k(\vec{r}_2)\rangle_{\omega} \exp[-i\omega(t-t')], \quad (A1)$$

$$\langle x_k(\vec{r}_2,t') x_i(\vec{r}_1,t)\rangle = \int_{-\infty}^{\infty} \frac{d\omega}{2\pi} \langle x_k(\vec{r}_1) x_i(\vec{r}_2)\rangle_{-\omega} \exp[-i\omega(t-t')], \quad (A2)$$

where the unsimmetrized spectral power densities are connected as follows [26]

$$\langle x_k(\vec{r}_1) x_i(\vec{r}_2)\rangle_{-\omega} = \langle x_i(\vec{r}_1) x_k(\vec{r}_2)\rangle_{\omega} \exp[-\hbar\omega/k_B T]. \quad (A3)$$

The symmetrized SPD is equal

$$\left(x_i(\vec{r}_1) x_k(\vec{r}_2)\right)_{\omega} = (1/2)\left\{\langle x_i(\vec{r}_1) x_k(\vec{r}_2)\rangle_{\omega} + \langle x_k(\vec{r}_1) x_i(\vec{r}_2)\rangle_{-\omega}\right\}. \quad (A4)$$

In accordance with the fluctuation dissipation theorem

$$\left(x_i(\vec{r}') x_k(\vec{r}'')\right)_{\omega} = \hbar \mathrm{Coth}(\hbar\omega/2k_B T)(\chi_{ik}(\vec{r}',\vec{r}'';\omega) - \chi_{ki}^{*}(\vec{r}',\vec{r}'';\omega))/2i, \quad (A5)$$

where

$$\chi_{ik}(\omega) = \frac{i}{\hbar}\int_0^{\infty} d(t-t') \exp[i\omega(t-t')] \langle x_i(t) x_k(t') - x_k(t') x_i(t)\rangle, \quad (A6)$$

is the Fourier transform of the linear response tensor.

Let's write the continuity equation for two spatial points

$$i\omega'\rho(\vec{r}',\omega') = \partial j_\alpha(\vec{r}',\omega')/\partial x'_\alpha,$$
$$i\omega''\rho(\vec{r}'',\omega'') = \partial j_\beta(\vec{r}'',\omega'')/\partial x''_\beta \quad . \tag{A7}$$

Taking into account that for stationary processes we have

$$\overline{x_k(\omega')x_i(\omega'')} = 2\pi \langle x_i x_k \rangle_{\omega'} \delta(\omega'+\omega''), \tag{A8}$$

it is not difficult to verify that

$$\omega^2 \langle \rho(\vec{r}')\rho(\vec{r}'') \rangle_\omega = \frac{\partial^2}{\partial x'_\alpha \partial x''_\beta} \langle j_\alpha(\vec{r}') j_\beta(\vec{r}'') \rangle_\omega, \tag{A9}$$

$$i\omega \langle \rho(\vec{r}') j_\beta(\vec{r}'') \rangle_\omega = \frac{\partial}{\partial x'_\alpha} \langle j_\alpha(\vec{r}') j_\beta(\vec{r}'') \rangle_\omega, \quad i\omega \langle j_\alpha(\vec{r}')\rho(\vec{r}'') \rangle_\omega = \frac{\partial}{\partial x''_\beta} \langle j_\alpha(\vec{r}') j_\beta(\vec{r}'') \rangle_\omega, \tag{A10}$$

$$-i\omega \langle \rho(\vec{r}')\rho(\vec{r}'') \rangle_\omega = \frac{\partial}{\partial x'_\alpha} \langle j_\alpha(\vec{r}')\rho(\vec{r}'') \rangle_\omega = \frac{\partial}{\partial x''_\beta} \langle \rho(\vec{r}') j_\beta(\vec{r}'') \rangle_\omega, \tag{A11}$$

$$-i\omega \langle \rho(\vec{r}'')\rho(\vec{r}') \rangle_\omega = \frac{\partial}{\partial x'_\alpha} \langle \rho(\vec{r}'') j_\alpha(\vec{r}') \rangle_\omega = \frac{\partial}{\partial x''_\beta} \langle j_\beta(\vec{r}'')\rho(\vec{r}') \rangle_\omega. \tag{A12}$$

## APPENDIX B: DERIVATION OF SPD $g_{ZZ}^{(e)}(\vec{r}_1,\vec{r}_2;\omega)$ IN A HYDRODYNAMICAL APPROACH

Schaich [27] derived two expressions for the Fourier transform $\chi(z_1,z_2;Q;\omega)$ with respect to lateral coordinates as follows

$$\chi(\vec{r}_1,\vec{r}_2;\omega) = \iint \frac{d\vec{Q}}{(2\pi)^2} \exp[i\vec{Q}(\vec{r}'_\| - \vec{r}''_\|)] \chi(z_1,z_2;Q;\omega), \tag{B1}$$

where we use here the expression in Eq.(71) from the cited paper

$$\chi(z_1,z_2;Q;\omega) = -\frac{\omega_P^2}{4\pi\beta^2} \left\{ -\delta(z'-z'') + \frac{Q_L^2 - Q^2}{2Q_L} \times \right.$$
$$\left. \times \left[ \exp(-Q_L |z'-z''|) + (1+\gamma)\exp(Q_L z' + Q_L z'') \right] \right\}, \quad z',z'' < 0 \tag{B2}$$

where $\omega_P$ is the plasma frequency, $\beta^2 = (3/5)v_F^2$ is the nonlocal parameter and we have taken into account our definition of the linear response function in Eq.(A6) and different sign of the Kubo formula in [26] and [29].

Integration in (B1) with respect to the angle variable yields

$$\chi(\vec{r}_1,\vec{r}_2;\omega) = (2\pi)^{-1} \int_0^\infty Q dQ\, J_0(Q|\vec{r}'_1 - \vec{r}''_2|)\, \chi(z'_1,z''_2;Q;\omega). \tag{B3}$$

Taking into account Eq.(B3) in Eq.(9) we obtain

$$g_{ik}^{(E)}(\vec{r}_1,\vec{r}_2;\omega) = \frac{\hbar\omega_P^2}{4\pi^2\beta^2[\exp(\hbar\omega/k_BT)-1]} \text{Im}\left\{\int_0^\infty QdQ\left(\frac{Q_L^2-Q^2}{2Q_L}\right)\iint_V d^3r'd^3r''\right.$$

$$\left.\frac{(h+z')(h+z'')J_0(Q|\vec{r}_1'-\vec{r}_2''|)[\exp(-Q_L|z'-z''|)+(1+\gamma)\exp(-Q_Lz'-Q_Lz'')]}{\left[(r_\|')^2+(h+z')^2\right]^{3/2}\left[(r_\|'')^2+(h+z'')^2\right]^{3/2}}\right\}, \quad \text{(B4)}$$

where we have changed a sign of variables $z', z''$, and an integrations over the volume means in case of a half-space

$$\iint_V d^3r'd^3r''... = \int_0^\infty dz'\int_0^\infty dz''\int_0^\infty r_\|' dr_\|'\int_0^\infty r_\|'' dr_\|''\int_0^{2\pi} d\phi_1\int_0^{2\pi} d\phi_2.... \quad \text{(B5)}$$

In case of a film we should integrating over a thickness of the film.

Elementary integration in Eq.(B4) yields in Eq.(12). For completeness we also give all expressions used in Eq.(B4),(12)

$$Q_L^2 = Q^2 + (\omega_P^2 - \omega^2 - i\omega\nu)/\beta^2, \quad Q_T^2 = Q^2 - (\omega^2/c^2)\varepsilon(\omega),$$
$$Q_T^{02} = Q^2 - (\omega^2/c^2), \quad \varepsilon(\omega) = 1 - \omega_P^2/(\omega^2 + i\omega\nu), \quad \text{(B6)}$$
$$\gamma = 2Q^2[1-\varepsilon(\omega)]/\{Q_L[\varepsilon(\omega)Q_T^0 + Q_T]+[\varepsilon(\omega)-1]Q^2\}$$

It should be noted that using another formula for $\chi$ from [27] yields to similar results.

## APPENDIX C: THE KOHN-SHAM SYSTEM OF EQUATIONS

Owing to the geometry of the problem under consideration (a half-space, a plane-parallel film) the one particle electron wave function is factorized yielding to the one dimensional Schrodinger equation with an effective potential. This effective potential and electron's density nearby a sample surface are sought for self-consistently. Thus, the wave function is written [32] as follows

$$\psi_\nu(\vec{r}) = S^{-1/2} \exp(i\vec{Q}\vec{\rho})\phi_\ell(z) , \quad \text{(C1)}$$

where $\nu = \{Q,\ell\}$. The electron density is

$$n(\vec{r}) = \sum_\nu f_\nu |\psi_\nu(\vec{r})|^2 , \quad \text{(C2)}$$

where

$$f_\nu = [\exp(E_\nu - \mu)/k_BT + 1]^{-1}, \quad E_\nu = \hbar^2Q^2/2m + \varepsilon_\ell, \quad \text{(C3)}$$

$\mu$ is the chemical potential.

Substitution of (C1), (C3) into (C2) followed by an integration instead of summation with respect to the lateral wavenumber $Q$ yields

$$n(\vec{r}) = \frac{mk_BT}{\pi\hbar^2}\sum_\ell\left\{\frac{\mu-\varepsilon_\ell}{k_BT} + \ln\left[1+\exp\left(\frac{\varepsilon_\ell-\mu}{k_BT}\right)\right]\right\}|\phi_\ell(z)|^2 , \quad \text{(C4)}$$

which tends to

$$n_0(\vec{r}) = \frac{m}{\pi\hbar^2}\sum_\ell \theta(E_F - \varepsilon_\ell)|\phi_\ell(z)|^2 (E_F - \varepsilon_\ell), \tag{C5}$$

at $T \to 0$, where $E_F \equiv \mu(T=0)$.

The self-consistent procedure includes Eq.(C4) and the Schrodinger equation

$$\left[-(\hbar^2/2m)d^2/dz^2 + V_{eff}(z)\right]\phi_\ell(z) = \varepsilon_\ell \phi_\ell(z), \tag{C6}$$

$$V_{eff}(z) = e^2 \iiint d\vec{r}' \frac{[n(\vec{r}') - n_+]}{|\vec{r} - \vec{r}'|} + V_{XC}(\vec{r}), \tag{C7}$$

where $n_+$ is the positive jellium background, $V_{XC}$ is the exchange-correlation potential, which we did not take into account in this our paper, and the neutrality condition

$$n_+ d = \frac{mk_B T}{\pi\hbar^2}\sum_\ell \left\{\frac{\mu - \varepsilon_\ell}{k_B T} + \ln\left[1 + \exp\left(\frac{\varepsilon_\ell - \mu}{k_B T}\right)\right]\right\}. \tag{C8}$$

This is equation for finding the chemical potential.

It should be noted that the Hartree term in Eq.(C7) is divergent due to infinite extension of a sample along lateral directions, that is why take it in the following form

$$e^2 \iiint d\vec{r}' \frac{[n(\vec{r}') - n_+]}{|\vec{r} - \vec{r}'|} = \lim_{L\to\infty} e^2 \int_{-L}^{L}\int_{-L}^{L}\int_0^d \frac{[n(z') - n_+]dx'dy'dz'}{\sqrt{(x-x')^2 + (y-y')^2 + (z-z')^2}} =$$
$$= -2\pi e^2 \int_0^d dz'[n(z') - n_+]|z - z'| + (L\to\infty) \times 2\pi e^2 \int_0^d dz'[n(z') - n_+]. \tag{C9}$$

The self-consistent solution yields to the electro neutrality condition

$$\int_0^d dz'[n(z') - n_+] = 0. \tag{C10}$$

We follow the matrix way of solution described in Ref.[32-34] introducing the representation

$$\phi_\ell(z) = (2/d)^{1/2}\sum_{s=1}^\infty b_s^{(\ell)} \sin(s\pi z/d), \tag{C11}$$

accepting a zeroing of the wave function at some distance $3\pi/8k_F$ in a vacuum side, where $k_F$ is the Fermi wavenumber. The Schrodinger Eq.(C4) is solved in a matrix form

$$\frac{\hbar^2}{2m}\sum_{s=1}^\infty b_s^{(\ell)}\left(\frac{s\pi}{2}\right)^2 \delta_{ps} + \sum_{s=1}^\infty b_s^{(\ell)} V_{sp}^{eff} = \varepsilon_\ell b_p^{(\ell)}, \tag{C12}$$

which is the equation for eigenvalues and eigenfunctions.

We do not show a figure for the electron density distribution because of a majority of such figures in a literature; see, for example, in [32-35].

Taking self-consistently the eigenvalues and eigenfunctions of the electron's subsystem from the set of above described equations, we calculate the linear response function of the bounded system in accordance with Eq.(13), where the reduced response function $\chi^0(z_1, z_2; Q; \omega)$ may be written [33, 34] for noninteracting Kohn-Sham states as follows

$$\chi^0(z', z''; Q, \omega) = e^2 \sum_{\ell=1}^{\ell_M} \sum_{\ell'=1}^{\ell'_M} F_{\ell\ell'}(Q, \omega) \phi_\ell(z') \phi_{\ell'}(z') \phi_\ell(z'') \phi_{\ell'}(z''), \qquad (C13)$$

where $\phi_\ell(z)$ from Eq.(C11) with self-consistent coefficients $b_{\ell s}$,

$$\begin{aligned}F_{\ell\ell'}(Q,\omega) = &-(m^2/\pi\hbar^3 Q^2)\{2a_{\ell\ell'}(Q) + \\ &+ i\sqrt{\hbar^2 Q^2 k_\ell^2/m^2 - [a_{\ell\ell'}(Q) - \omega - i\eta]^2} - \\ &- i\sqrt{\hbar^2 Q^2 k_\ell^2/m^2 - [a_{\ell\ell'}(Q) + \omega + i\eta]^2}\}\end{aligned} \qquad (C14)$$

where $a_{\ell\ell'}(Q) = \hbar Q^2/2m - (\varepsilon_\ell - \varepsilon_{\ell'})/\hbar$ and $k_\ell^2 = (2m/\hbar^2)(E_F - \varepsilon_\ell)$, $\eta \to 0$.

Here in Eq.(C13) we have taken into account our definition in Eq.(10) differing from corresponding definition in [33].

**FIGURE CAPTIONS**

**Fig.1.** Normalized spectral power density of z-component of the electric thermal field $\tilde{g}_{zz}^{(E)}(h)$ versus a distance *h* in accordance with Eq.(12) (red curve), and in accordance with a phenomenological theory of thermally stimulated fields [1-3,10-14] (black line) at the fixed

frequency $\omega = 0.1\omega_P$. The functions are normalized at the value $g_{zz}^{(E)}(h=0)$. The quantity $h=0$ is fixed at the edge of the positive background of a metal.

**Fig.2.** Normalized value $g_{EE}(\omega)/u_{BB}(\omega)$ within a metal half-space ($h<0$) and $g_{EE}(\omega,h)/u_{BB}(\omega)$ in vacuum ($h>0$) at different fixed frequencies $\omega = 0.01\omega_P$, $\omega = 0.05\omega_P$, $\omega = 0.1\omega_P$ from the top to the bottom curves. Normalization was done to the spectral energy density $u_{BB}(\omega)$ of the black body radiation in vacuum in accordance with the Planck law.

**Fig.3.** Distance dependence of the normalized spectral power density $\tilde{g}_{zz}^{(E)}(h) = 2\tilde{g}_{xx,yy}^{(E)}(h)$ in accordance with Eq.(14) (red curves), and in accordance with a phenomenological theory of thermally stimulated fields [1-3,10-14] (black line) at the fixed frequency $\omega = 0.1\omega_P$. The red curve 1 corresponds to the case of IBM model (see details in the text). The case of self-consistent calculations of SPD with use of both terms $G^0(Q,\omega)$ and $G(Q,\omega)$ in Eq.(14) in the random phase approach (RPA) is demonstrated by the red curve 2. All curves are normalized at $g_{zz}^{(E)}(h=0)$ of the second red curve.